\documentclass[aps,12pt,prb]{revtex4}
\usepackage{enumerate}
\usepackage{amsmath}

\begin{document}
\baselineskip18pt

\title{Information dynamics in quantum theory}
\author{Piotr Garbaczewski\thanks{electronic address pgar@uni.opole.pl}}
\affiliation{Institute of Physics, University of Opole, 45-052
Opole, Poland}
\begin{abstract}
Shannon entropy  and Fisher information functionals are  known to quantify certain information-theoretic properties of
 continuous probability distributions of various  origins.
We carry out a systematic  study of these   functionals, while  assuming that the  pertinent  probability density
has a quantum mechanical appearance $\rho \doteq |\psi |^2$, with $\psi \in L^2(R)$. Their behavior in time,  due to
the  quantum   Schr\"{o}dinger  picture evolution-induced dynamics
of $\rho (x,t)$ is investigated as well, with an emphasis on thermodynamical features of quantum motion.
\end{abstract}
\maketitle
\vskip0.5cm
\section{Information-is there anything where nobody looks ?}

Shannon and   von Neumann entropies are typical information
 theory   tools  which can be used   to quantify  the information
content and possibly information loss/gain  incurred by  a quantum
system, initially   prepared in a specified  (macro)state.
 The von Neumann entropy   is known to  vanish on
 pure states, hence  one presumes to have   a complete information about
such state.  An incomplete information   concept     thus  seems to be  related  only  to mixed states.

    On the other  hand,  right  in connection with   pure
states of a quantum mechanical system, the Shannon  entropy  is known to  give an
  access to another information theory  level. Namely, it enables   quantifying   an information content
of  continuous  probability distributions, that can be  inferred
from  any   $\psi \in L^2(R^n)$  vector by means of the Born recipe
 $\rho \doteq |\psi |^2$.

 Since, in  physics,    the very concept of  entropy is  typically interpreted
 as  a measure of the degree of randomness    and the  tendency (trend)  of physical
systems to become less organized (disordered), it is   quite  natural to
think of entropy as about the measure of uncertainty or disorder.
Notions of information and uncertainty are deeply intertwined.  Spectacular examples of this intertwine are
provided by   information theory measures employed in quantum theory in the description of so-called entropic
uncertainty relations that are valid in $L^2(R^n)$ for Born postulate-induced continuous
probability densities, \cite{mycielski}.

The term \it information, \rm in the
present context,  may be  literally understood as the \it inverse of
uncertainty\rm.   As far as the notion of  an  \it  organization  \rm is concerned, the
 Shannon entropy  (and a number of  other entropic measures)  is known to   quantify
   the degree of  the probability distribution   complexity, \cite{panos},  and
 (de)localization, \cite{gar}, for stationary and non-stationary
Schr\"{o}dinger wave packets.

Let us tentatively accept the casual statement that a (thermodynamically)  \it isolated \rm system is represented in quantum mechanics
 by a state vector  which
conveys statistic predictions for measurement outcomes.  Then, we are tempted to  identify and quantify an
information  content of  a  state vector, even   though we   know that the  von  Neumann entropy
(the standard quantum measure of information)  identically vanishes  on a pure quantum state.  The related  semantic word-game due
to Roger  Penrose is worth mentioning: "when a system has a state $|\psi \rangle$  there ought to be some property in the system
that corresponds  to its $|\psi \rangle$-ness", \cite{penrose}.

A more formal issue appears if we  pass to  the quantum dynamics, when we  should in principle  address an information theoretic
interpretation of quantum evolutions.
Clearly,  in terms of the von Neumann entropy, nothing illuminating
  can be said about the quantum motion of pure or mixed  states  of an isolated system, since  the
unitary evolution leaves the von Neumann entropy intact. Therefore,  the von Neumann entropy behavior in time  may  become
interesting only  if we pass from isolated to  (thermodynamically) \it  open \rm  systems.

Quite to the contrary, the Shannon entropy of a continuous  probability  distribution  may show up   a  non-trivial
pattern of temporal behavior which deserves a closer inspection.  Even, if our attention  is confined  to  an   \it  isolated
\rm quantum system in  its   pure state.

\section{Information functionals and indeterminacy relations}

Given  an $L^2(R)$-normalized function  $\psi (x)$. We denote $({\cal{F}}\psi )(p)$ its Fourier
transform. The corresponding probability densities follow:
 $\rho (x) = |\psi (x)|^2$ and $\tilde{\rho }(p) = |({\cal{F}}\psi )(p)|^2$.\\
We introduce the related position and momentum  information  (differential, e.g. Shannon) entropies:
\begin{equation}
{\cal{S}}(\rho )\doteq S_q =  - \langle \ln \rho \rangle = - \int \rho (x) \ln \rho (x)  dx
\end{equation}
    and
    \begin{equation}
    {\cal{S}}(\tilde{\rho })\doteq
S_p= - \langle  \ln \tilde{\rho }\rangle =   - \int \tilde{\rho }(p) \ln \tilde{\rho }(p)  dp
\end{equation}
 where  ${\cal{S}}$ denotes the  Shannon
 entropy for a continuous probability  distribution.  For the sake of clarity, we
 use dimensionless quantities,  see e.g. however\cite{gar1} for a discussion of how to handle dimensional quantities in the Shannon entropy
 definition.

We assume both entropies to take finite values. Then,
there holds the familiar   entropic uncertainty relation \cite{mycielski}:
\begin{equation}
S_q + S_p \geq  (1 + \ln \pi ) \, .  \label{uncertainty}
\end{equation}

If  following  conventions we define
the squared  standard  deviation   value  for an observable $A$ in a pure
state $\psi $ as
$(\Delta A)^2 = (\psi , [A - \langle A\rangle ]^2 \psi )$ with $\langle A
\rangle  = (\psi , A\psi)$, then for the  position $X$ and momentum $P$ operators
we have the following version of the  entropic uncertainty relation  (here expressed
through so-called entropy powers, see e.g.   \cite{petz}, $\hbar \equiv 1$):
\begin{equation}
\Delta X \cdot \Delta P \geq  {\frac{1}{2\pi e}} \,
 \exp[{\cal{S}}(\rho )  + {\cal{S}}(\tilde{\rho })] \geq {\frac{1}{2}} \,  \label{un}
\end{equation}
which is  an alternative version of the entropic uncertainty relation.

An important property of the Shannon entropy  ${\cal{S}}(\rho ) $ is that for a continuous
 probability distribution $\rho (x)$ with an  arbitrary  finite mean   $\langle X\rangle $ and  a  fixed variance
 $\sigma ^2 = \langle (X- \langle X\rangle )^2\rangle = \Delta X ^2 $ we would have
\begin{equation}
{\cal{S}}(\rho )\leq  {\frac{1}2}  \ln (2\pi e \sigma ^2) \, .
\end{equation}
${\cal{S}}(\rho )$ becomes maximized  in the set of such densities  if and only if $\rho $
is a Gaussian  with  variance $\sigma ^2$.  For Gaussian densities,   $ (2\pi e )\Delta X \cdot \Delta P =
 \exp[{\cal{S}}(\rho )  + {\cal{S}}(\tilde{\rho })]$ holds true, but the     minimum $1/2$
  on the right-hand-side of Eq.~(\ref{un}),   is  not necessarily  reached.

  In below, we shall devote some attention to  the Fisher  information measure\cite{gar,stam}:
\begin{equation}
{\cal{F}} (\rho ) \doteq  \langle ({\nabla \ln \rho })^2\rangle = \int {\frac{(\nabla \rho )^2}{\rho }}  dx
\end{equation}
which stays in a remarkable   relationship with the Shannon entropy of the very same continuous  probability distribution\cite{stam}:
\begin{equation}
{\cal{F}}(\rho ) \geq (2\pi e)  \exp [-2{\cal{S}}(\rho )] \geq  {\frac{1}{\sigma ^2}} \, . \label{chain}
\end{equation}
Clearly, we have ${\cal{F}}(\rho ) \geq (1/\sigma ^2)$ with the equality allowed only if $\rho $ is a Gaussian with variance $\sigma
 ^2$.

 Let us notice that in view of properties of  the Fourier transform, there is a complete symmetry between  the
 inferred  information-theory
 functionals. After the Fourier transformation, the Parceval identity implies that  the chain of
  inequalities  Eq.~(\ref{chain}) can be  faithfully reproduced (while replacing $\rho $  by $\tilde{\rho }$) for
 the "momentum -space" density $\tilde{\rho }$ with the  variance $\tilde{\sigma }^2$.  As a consequence, taking into   account
 the entropic uncertainty relation Eq.~(\ref{uncertainty}), we arrive at\cite{mycielski}:
\begin{equation}
4 \tilde{\sigma }^2 \geq  2(e\pi )^{-1}  \exp[-2  \langle \ln \tilde{\rho }\rangle ] \geq  (2e\pi ) \exp[ 2 \langle \ln  \rho \rangle ]
\geq \sigma ^{-2} \label{chain1}
\end{equation}

Let us consider a momentum operator $P$ that is conjugate  to the position operator  $X$ in the adopted dimensional convention $\hbar \equiv 1$.
Setting $P= - i d/dx$ and presuming that all  averages are finite,  we get:
\begin{equation}
[\langle P^2\rangle - \langle P\rangle ^2] =  (\Delta P)^2= \tilde{\sigma }^2 \, .
\end{equation}
The  standard indeterminacy relationship
 $\sigma  \cdot \tilde{\sigma }\geq (1/2)$  follows.

  In the above, no explicit time-dependence
has been  indicated, but all derivations go through with
any wave-packet  solution $\psi (x,t)$ of the Schr\"{o}dinger equation. The  induced dynamics of
 probability densities may imply the  time-evolution of  entropies: $S_q(t), S_p(t)$ and thence the dynamics of
 quantum  uncertainty measures  $ \Delta X  (t) =  \sigma (t)$ and $ \Delta P(t)= \tilde{\sigma }(t) $.

\section{Hydrodynamical velocity fields and  their variances}

Let us consider the Schr\"{o}dinger equation:
\begin{equation}
i\partial _t  \psi  = - D  \Delta  \psi   +
{\frac{{\cal{V}}}{2mD}} \label{Schroedinger} \psi \, .
\end{equation}
where the  potential ${\cal{V}}= {\cal{V}}(\overrightarrow{x},t)$ (possibly time-dependent)  is  a  continuous
(it is useful, if bounded from below)  function  with dimensions of
 energy, $D=\hbar /2m$.

By employing the  Madelung decomposition:
\begin{equation}
 \psi = \rho ^{1/2} \exp(is/2D)
 \end{equation}
   with the
phase function $s=s(x,t)$  defining $v=\nabla s $ we readily  arrive at the continuity equation
\begin{equation}
\partial _t \rho = - \nabla (v \rho )
\end{equation}
 and the generalized Hamilton-Jacobi equation:
\begin{equation}
\partial _ts +\frac{1}2 ({\nabla }s)^2 + (\Omega -  Q) = 0 \label{jacobi1}
\end{equation}
where, after  introducing an additional velocity field
\begin{equation}
u(x,t) = D\nabla \ln \rho (x,t) \, ,
\end{equation}
we have
\begin{equation}
Q  =  2D^{2}{\frac{\Delta \rho ^{1/2}} {\rho ^{1/2}}} = {\frac{1}2}
u^2 + D \nabla \cdot u \, .
\end{equation}

If  a quantum mechanical expectation value of the standard Schr\"{o}dinger  Hamiltonian  $\hat{H}= -(\hbar ^2/2m) \Delta + V$
  exists (i.e.  is finite\cite{gar2}),
  \begin{equation}
 \langle \psi | \hat{H}|\psi \rangle \doteq E < \infty
 \end{equation}
  then the unitary  quantum dynamics warrants that this value is a  constant
 of the Schr\"{o}dinger picture evolution:
\begin{equation}
{\cal{H}} = {\frac{1}2} [\left< {v}^2\right> + \left< {u}^2\right>]  +
\left<\Omega \right>  =  - \left< \partial _t s\right>  \doteq {\cal{E}} = {\frac{E}m}= const  \, . \label{total}
\end{equation}
Let  us notice that  $\langle u^2\rangle = - D \langle \nabla u \rangle $ and therefore:
\begin{equation}
{\frac{D^2}{2}} {\cal{F}} =   {\frac{D^2}{2}} \int {\frac{1}{\rho }}
\left({\frac{\partial \rho }{\partial x }} \right)^2\, dx  =   \int
\rho \cdot  {\frac{u^2}{2}} dx = -   \langle Q \rangle \, .
\label{Fisher1}
  \end{equation}

Let us observe that  $D^2{\cal{F}}$  stands for  the  mean square deviation  value  of a function $u(x,t)$ about its mean value
$\langle u \rangle =0$, whose vanishing is a consequence of the boundary conditions (here, at infinity):
\begin{equation}
(\Delta u)^2 \doteq \sigma _u^2 =  \langle [u- \langle u\rangle ]^2\rangle = \langle u^2\rangle  = D^2 {\cal{F}} \, .
\end{equation}
The  mean square deviation of $v(x,t)$ about its mean value $\langle v\rangle$ reads:
\begin{equation}
(\Delta v)^2 \doteq \sigma ^2_v = \langle v^2 \rangle -
\langle v \rangle ^2\, .
\end{equation}
It is clear, that with the definition $P= -i(2mD) d/dx$, the mean value of the operator $P$ is related to the mean value of a
function $v(x,t)$  (we do not discriminate between  technically  different implementations  of the mean):  $\langle P\rangle = m\langle v \rangle $.
Accordingly,
\begin{equation}
\tilde{\sigma }^2  = (\Delta P)^2 = \langle P^2\rangle - \langle P\rangle ^2
\end{equation}

Moreover, we can directly check that with $\rho = |\psi |^2$ there
holds\cite{hall}:
\begin{equation}
{\cal{F}}(\rho ) = {\frac{1}{D^2}} \sigma ^2_u = \int dx |\psi |^2[\psi '(x)/\psi (x) + {\psi ^*}'(x) /\psi ^*(x)]^2=   \label{ha}
\end{equation}
$$
4\int dx {\psi '}^*(x) {\psi '}(x)  + \int dx |\psi (x)|^2 [ \psi '(x)/\psi (x) -{\psi ^*}'(x) /\psi ^*(x)]^2  =
$$
$$
{\frac{1}{m^2D^2}} [ \langle P^2\rangle - m^2 \langle v^2 \rangle ] = {\frac{1}{m^2D^2}}[ (\Delta P)^2 -  m^2 \sigma ^2_v]
$$
i.e.
\begin{equation}
 m^2(\sigma ^2_u + \sigma ^2_v) = \tilde{\sigma }^2 \, .
\end{equation}
  It is interesting to notice that $\langle (P - mv )\rangle  =0$
and the corresponding mean square deviation  reads:
\begin{equation}
 \langle (P-mv)^2\rangle = \langle P^2\rangle - m^2\langle v^2\rangle = m^2D^2 {\cal{F}} \, .
 \end{equation}

An interesting outcome of this discussion is a definite sharpening  of an  upper bound in the inequalities Eqs.~(\ref{chain}). Namely,
by passing to dimensionless  quantities in Eqs.~(\ref{ha}) (e.g. $2mD\equiv 1$), and denoting $p_{cl} \doteq
 (\arg \,  \psi (x,t) )' $ we get:
\begin{equation}
{\cal{F}} = 4[\langle P^2\rangle - \langle p^2_{cl}\rangle ] = 4[(\Delta P)^2 -  (\Delta p_{cl})^2] = 4[\tilde{\sigma }^2 - \tilde{\sigma }_{cl}^2]
\end{equation}
and therefore  the chain of inequalities Eq.~(\ref{chain}) gets a sharper form with a manifest upper bound  for the Shannon entropy
 of $\rho = |\psi |^2$ set by:
 \begin{equation}
4\tilde {\sigma }^2 \geq 4[\tilde{\sigma }^2 - \tilde{\sigma }_{cl}^2] = {\cal{F}}      \geq (2\pi e)  \exp [-2{\cal{S}}(\rho )]
\geq  {\frac{1}{\sigma ^2}} \, .
\end{equation}
We recall that all "tilde" quantities can be  deduced from the  once  given $\psi $ and  its Fourier transform $\tilde{\psi}$.

\section{Thermodynamical features of the quantum dynamics}

We have emphasized that  a pure state of the quantum theory  and its  Schr\"{o}dinger picture dynamics
 are normally attributed to a thermodynamically \it isolated \rm quantum system.
 We would like to demonstrate that a number of
 essentially thermodynamical features is encoded in this innocent-looking, apparently    non-thermodynamical
   regime. To this end  some basic notions of the non-equilibrium thermodynamics must be introduced.

  \subsection{Quantum detour - thermodynamics of  open systems}

We  shall  give a concise  resume of the pertinent framework following\cite{alicki}.
It is taken for granted that in case of an open quantum system,  the bath drives a  system to an equilibrium state.
The state of the system plus reservoir  is described  by a  density matrix. Let
$\rho _t$  be  the reduced  density matrix of a quantum system in  a combined  weak coupling
and adiabatic approximation of the general system-reservoir dynamic problem, $t\geq 0$:
\begin{equation}
{\frac{d}{dt}} \rho _t = -i [H_{sys}(t), \rho _t ] + L_{diss}(t)\rho _t \doteq L(t) \rho _t
\end{equation}
We introduce the following thermodynamical notions:
(i)  an \it  internal energy \rm  of the system   $E(t) = Tr(\rho _tH_{sys}(t))$, (ii)  the
\it  work \rm    performed on the system by external forces $W(t) = \int_0^t Tr[\rho _s ({\frac{d}{ds}}
H_{sys}(s))] ds $. (iii) the \it  heat \rm supplied to the system by the reservoir
$Q(t) = \int_0^t Tr[({\frac{d}{ds}}\rho _s) H_{sys}(s)] ds$.

The laws of thermodynamics, tailored to the manifestly non-equilibrium dynamical regime  can now be formulated.
 The first law of thermodynamics reads:
\begin{equation}
{\frac{d}{dt}} E(t) = {\frac{d}{dt}}W(t)  + {\frac{d}{dt}} Q(t) \, .
\end{equation}
Let us introduce   the  relative entropy $(\rho|\sigma ) = Tr(\rho \ln \rho - \rho \ln \sigma )$ and
the  account for the  stationary
state input $L(t)\rho _{eq} =0$ , with $\rho _{eq} = Z^{-1} \exp[-\beta H_{sys}(t)]$.
 Then, the second law of thermodynamics takes the form:
 \begin{equation}
 {\frac{d}{dt}} S(\rho _t|\rho _{eq}) = \sigma (\rho _t) + {\frac{1}T}{\frac{dQ}{dt}}
\end{equation}
where $\sigma (\rho _t)\geq 0$  is called   the  entropy production, while $\dot{Q}/T$
refers to the entropy/heat exchange with the bath. Obviously, we have   $TdS \geq dQ$.

\subsection{Back to classical non-equilibrium thermodynamics}

For the record,  we indicate that the following  hierarchy of
thermodynamic systems is adopted  in the present
paper\cite{glansdorf,kondepudi}: {\it isolated} with no energy and
matter exchange with the environment,  {\it closed} with the
energy   but {\it no } matter exchange and
 {\it open} where energy-matter exchange  is unrestricted.

 Our previous discussion was confined to an open quantum system.
Accepting the standard text-book wisdom that all isolated systems evolve to the state of equilibrium
 in which  the entropy reaches its maximal value, we shall pay attention to
  {\it  closed} random  systems and their somewhat different   asymptotic properties.
A concise resume of a non-equilibrium thermodynamics of {\it closed}
systems comprises the  $I^{st}$  law of thermodynamics
\begin{equation}
\dot{U} = \dot{Q} + \dot{W}
\end{equation}
and  the $II^{nd}$  law of thermodynamics:
\begin{equation}
\dot{S}= \dot{S}_{int} + \dot{S}_{ext}\, ,
\end{equation}
where $\dot{S}_{int}\geq 0$ and $\dot{S}_{ext}  = \dot{Q}/T$, c.f.
\cite{glansdorf,kondepudi}.   Let us emphasize that $\dot{Q}$ and
$\dot{W}$ are  always well defined, but the adopted (time derivative) notation does
not imply that  one may infer $Q$ and $W$ as legitimate
thermodynamic functions, c.f. an issue of "imperfect differentials"
 in classical thermodynamics.

Thermodynamic extremum principles are usually invoked in
connection with the large time  behavior of  irreversible
processes. One  looks for direct realizations of the entropy
growth paradigm, undoubtedly valid for  isolated systems,
\cite{mackey}. Among a number of standard thermodynamic
extremum principles, we   recall  a specific one  named the
Helmholtz extremum principle. If the temperature $T$ and  the
available volume $V$ are kept constant, then the minimum of the
Helmholtz free energy
\begin{equation}
 F= U - TS
 \end{equation}
   is preferred  in the course of the system evolution in time,
 and  there holds\cite{kondepudi}
\begin{equation}
 \dot{F} = - T\dot{S}_{int} \leq 0
\end{equation}
In below, we shall analyze the validity of thermodynamic principles
  and the role played by the direct analog of the Helmholtz free energy,
 in case of  quantum motion,  and specifically in the
seemingly non-thermodynamical context of the Schr\"{o}dinger picture evolution.

\subsection{Thermodynamical features  of the quantum motion - closed systems in action}

We come back to  the Schr\'{o}dinger picture evolution of pure states in $L^2(R)$.
We impose  the natural boundary data on quantum motion and they are implicit
(vanishing of various expressions at integration boundaries) in all averaging procedures in below.
One must be aware that we pass-by a number of mathematical subtleties and take for granted that
various computational steps are allowed.

The continuity equation is a direct consequence of the Schr\"{o}dinger equation. It is less obvious
that, after employing the hydrodynamical velocity fields $u(x,t)$ and $v(x,t)$, the Fokker Planck equation for
 $\rho = |\psi |^2$ may be deduced. We have:
\begin{equation}
\partial _t\rho =
D\triangle \rho -  \nabla \cdot ( b \rho )
\end{equation}
where $b=v+u =\nabla (s+D\ln \rho )$ where $u=D\nabla \ln
\rho $.

The Shannon entropy of a continuous probability distribution ${\cal{S}} = -\langle \ln \rho \rangle $ follows and yields
\begin{equation}
  D \dot{\cal{S}}  =  \left< {v}^2\right>
    -  \left\langle {b}\cdot {v} \right\rangle \doteq D (\dot{\cal{S}}_{int}    +  \dot{\cal{S}}_{ext})
\end{equation}
which is a straightforward analog of the  $II^{nd}$ law of thermodynamics in the considered  quantum mechanical context:
\begin{equation}
\dot{\cal{S}}_{int} = \dot{\cal{S}} -  \dot{\cal{S}}_{ext}  =   (1/D)\left< {v}^2\right> \geq 0
 \Rightarrow \dot{\cal{S}}\geq \dot{\cal{S}}_{ext} \, .
\end{equation}
 To address an analog of the  $I^{st}$ law we need to translate
to the present setting the previously discussed thermodynamic
notions of
$U$ and $F = U-TS$, where the notion of \it temperature \rm is the most serious obstacle. We have no obvious notion of temperature for
quantum systems in their pure states (for large molecules, like fullerenes or the likes, the notion of internal temperature makes
sense, but we aim to consider  any quantum system in a pure state, small or large).  Therefore, we shall  invoke a dimensional
artifice\cite{broglie}.

We formally introduce
\begin{equation}
  k_BT_0\doteq \hbar \omega _0\doteq mc^2
\end{equation}
   and thence
   \begin{equation}
D=\hbar/2m \equiv  k_BT_0/m\beta _0
\end{equation}
 with  $\beta _0 \equiv 2\omega _0 =
2mc^2/\hbar $, and so arrive at
\begin{equation}
 k_BT_0\dot{\cal{S}}_{ext} = \dot{Q}\, .
 \end{equation}

In view of:
\begin{equation}
v=\nabla s =  b- u =  \nabla (s + D\ln \rho ) - D \nabla \ln  \rho \doteq
\end{equation}
$$
 - {\frac{1}{m\beta }}\nabla (V + k_BT_0\ln \rho )\doteq - {\frac{1}{m\beta _0}}\nabla \Psi \, ,
$$
where  the time-dependent potential
\begin{equation}
V = V(x,t) \doteq  - m\beta _0(s + D\ln \rho )
\end{equation}
 is defined to stay in a  notational  conformity with the standard   Smoluchowski process
 (Brownian motion in a conservative force field\cite{gar}) definition $b=-\nabla V/m\beta _0 $,
 we finally  get
\begin{equation}
- m\beta  \langle s\rangle \equiv \langle \Psi \rangle =\langle V \rangle - T S  \Longrightarrow  F = U - T S  \, ,
\end{equation}
where $U= \langle V \rangle $ and $F=\langle \Psi \rangle $.

Remembering about an explicit time dependence of  $ b(x,t) =  -(1/m\beta _0)\nabla V(x,t)$, we  finally  arrive at
  the  $I^{st}$ law of thermodynamics in
the present quantum context:
\begin{equation}
\dot{U}=
\langle \partial _tV\rangle  - m\beta _0 \langle  b v \rangle =
\dot{W} + \dot{\cal{Q}}\, .
\end{equation}
The externally performed work  entry reads $\dot{W}=\langle \partial _tV\rangle $.
But:
$$V= - m\beta  s  -k_BT\ln \rho  \,  \Longrightarrow  \,
\langle \partial _t V\rangle  = -  m\beta _0\langle \partial _t s\rangle  =\dot{W}$$
and therefore
\begin{equation}
- {\frac{d}{dt}} \langle s \rangle = -  \langle v^2\rangle - \langle \partial _t s \rangle  \Rightarrow
\dot{F} =  - T_0 \dot{S}_{int} + \dot{W}
\end{equation}
where $\dot{S}_{int}\geq 0$.

If  a quantum mechanical expectation value of the standard Schr\"{o}dinger  Hamiltonian  $\hat{H}= -(\hbar ^2/2m) \Delta + V$
  exists (i.e.  is finite),
 $\langle \psi | \hat{H}|\psi \rangle \doteq E < \infty $, then
the unitary  quantum dynamics warrants that this value is a  constant
 of the Schr\"{o}dinger picture evolution which (c.f. Eq.(\ref{jacobi1})) implies:
\begin{equation}
{\cal{H}} = {\frac{1}2} [\left< {v}^2\right> + \left< {u}^2\right>]  +
\left<\Omega \right>  =  - \left< \partial _t s\right>  \doteq {\cal{E}} = {\frac{E}m}= const  \, . \label{total}
\end{equation}
Consequently, in the thermodynamical description of the  quantum motion, we encounter a  never vanishing constant  work term
\begin{equation}
  \dot{W} = m\beta _0{\cal{E}} = \beta _0\langle
 \hat{H}\rangle \, .
 \end{equation}
The associated Helmholtz-type  extremum principle reads:
\begin{equation}
\dot{F} - m\beta _0{\cal{E}}  = - T_0 \dot{S}_{int} \leq  0\, . \label{freeenergy}
\end{equation}
It is instructive to notice that
\begin{equation}
T\dot{S}_{int} = T\dot{S} - \dot{Q} \geq 0
 \Longleftrightarrow \dot{Q}\leq T\dot{S}
\end{equation}
 goes in parallel with
\begin{equation}
 \dot{F} \leq \dot{W} = \beta _0\langle \hat{H}\rangle \,  .
\end{equation}

Let us stress that the  non-vanishing  external work term is generic  to the quantum motion. If a stationary state is considered, our
$\langle \hat{H}\rangle $ is equal to  a corresponding energy eigenvalue.

 For negative eigenvalues,
the work term receives an interpretation  of the "work performed \it by \rm the system" (upon its, hitherto hypothetical,
 surrounding  ?). Then $\dot{F}$ is negative and $F$  may possibly have a chance to attain a minimum.

Since bounded from below Hamiltonians can  be replaced by positive operators, we may  in principle  view $m\beta _0{\cal{E}}=
\beta _0 \langle \hat{H}\rangle $ as a positive (constant and non-vanishing)  time rate of the
"work externally performed \it upon  \rm the system".   This observation encompasses   the case of positive energy spectra.
 Accordingly,  $\dot{F}$ may take both negative and  positive values. The latter   up to   an
 upper  bound $m\beta _0{\cal{E}}$.

 Basic features  of the non-equilibrium  thermodynamics of  \it closed \rm  irreversible   systems, somewhat surprisingly
  have been  reproduced in the  quantum Schr\"{o}dinger picture evolution.
  We  have identified  direct analogues of  the  $I^{st}$  and the $II^{nd}$   laws  of thermodynamics, together with the involved
 notions of $\dot{S}_{int} \geq 0$ and $\dot{S}_{ext} = (1/T) \dot{Q}$.

 An asymptotic behavior of the quantum motion is controlled by  the  analog of the $II^{nd}$ law:
\begin{equation}
  \dot{F} - \dot{W} = -m\beta _0{\frac{d}{dt}}(\langle s\rangle +{\cal{E}} t )=
  - T_0 \dot{S}_{int} \leq 0 \, .  \label{work}
\end{equation}
  where there appears an   work (performed \it upon \rm or  performed \it by \rm  the system)
   term $\dot{W} = \langle \partial _t V\rangle = m\beta _0 {\cal{E}} $ value  whose sign is indefinite
   (either positive or negative).

Let us notice that in classical non-equilibrium thermodynamics  the  so-called  minimum entropy production
 principle\cite{glansdorf} is often invoked in connection with the "speed" with which
  a minimum  of the Helmholtz free energy is approached.
  For sufficiently large times, when the system is in the vicinity of
  the stationary (equilibrium) state, one expects that  the the entropy production $T \dot{S}_{int} \geq 0$  is
   a monotonically   decaying function of time, i.e. that
  \begin{equation}
{\frac{d}{dt}} \dot{S}_{int} < 0 \, .
\end{equation}

The quantum motion looks different. In that case,  $\dot{F}$  may be positive and one cannot exclude
  transitions (including those of an oscillatory nature)  from negative to positive  $\dot{F}$ values and back.
It may
 happen that in certain quantum states, the Helmholtz free energy $F$ may have a minimum, a maximum, an
 infinite number of local minima and maxima, or none at all. There is no reason for the  minimum entropy production principle
 to be valid in quantum theory, except for very special cases.

 There is however a "speed" property which is special for the quantum case, with no dissipative counterpart.
  Namely, since the work term is
  a constant of quantum motion  and $\dot{F}+ T_0\dot{S}_{int} = m\beta _0{\cal{E}}$,  we have the
  following \it negative feedback \rm   relationship between the speeds
   of  the growth/decay of the entropy production and the  Helmholtz free energy  time rate:
 \begin{equation}
{\frac{d}{dt}} \dot{F} = -  T_0 {\frac{d}{dt}} \dot{S}_{int} \, .\label{feedback}
\end{equation}
If the Helmholtz free energy time rate drops down, the entropy production time rate
needs to increase  and in reverse.  Therefore a minimum  of $\dot{F}$ in principle  may  be achieved,
if  a maximum of the entropy production $\dot{S}_{int}$ is attained. In reverse, a maximum of $\dot{F}$ may arise
  in conjunction
with a minimum of $\dot{S}_{int}$.

Remembering that $T_0\dot{S}_{int} = m\beta _0 \langle v^2\rangle$ and exploiting the total mean energy formula,
 Eq. (\ref{total}), we can identify the respective "speeds":
 \begin{equation}
{\frac{d}{dt}} \dot{F} = \beta _0  {\frac{d}{dt}} (  m \langle u^2\rangle +  2\langle {\cal{V}}\rangle )
\end{equation}
and
\begin{equation}
T_0{\frac{d}{dt}} \dot{S}_{int}  =  m\beta _0{\frac{d}{dt}}
\langle v^2\rangle \,
\end{equation}
that stay in a feedback relationship. By recalling our discussion of Section
III, we realize that  variances of  the  hydrodynamical velocity fields decide about the time rate of the  entropy production and
Helmholtz free energy in the quantum case. They stay in the above mentioned  feedback relationship, consult e.g. also\cite{gar,gar3}.\\

{\bf  Acknowledgement:} The paper has been supported by the Polish
Ministry of Scientific Research and Information Technology under
the   grant No PBZ-MIN-008/P03/2003.\\


\begin{thebibliography}{99}
\bibitem{panos} K. Ch. Chatzisavvas , Ch. C. Moustakidis and C. P. Panos, J. Chem. Phys. {\bf 123}, 174111, (2005)
\bibitem{gar} P. Garbaczewski, J. Stat. Phys. {\bf 123}, 315, (2006)
\bibitem{mycielski} I. Bia{\l}ynicki-Birula and J.  Mycielski,  Commun. Math. Phys.  {\bf 44}, 129 (1975)
\bibitem{penrose}  R. Penrose, {\it The Emperor's New Mind}, Oxford University Press, Oxford,  1989
\bibitem{gar1} P. Garbaczewski,  Entropy, {\bf 7}[4], 253, (2005)
\bibitem{petz} Ohya, M. and Petz, D., {\it Quantum Entropy and Its use},  Springer-Verlag, Berlin, 1993
\bibitem{stam} A. J.  Stam, Inf. and  Control, {\bf 2}, 101,  (1959)
\bibitem{hall}  M. J. W. Hall, Phys. Rev. {\bf A 62}, 012107, (2000)
\bibitem{gar2} P. Garbaczewski, Rep. Math. Phys.   {\bf 56},  153, (2005)
\bibitem{alicki}  R. Alicki, in: \it Dynamics of Uncertainty \rm (eds. P. Garbaczewski and R. Olkiewicz), LNP vol.  597,
Springer-Verlag, Berlin,2002
\bibitem{glansdorf} P. Glansdorf and I. Prigogine, {\it Thermodynamic Theory of Structure, Stability
and Fluctuations},  (Wiley, NY, 1971)
\bibitem{kondepudi}  D. Kondepudi and I. Prigogine, {\it Modern Thermodynamics}, Wiley, NY, 1998
\bibitem{mackey} M. C. Mackey, M. Tyran-Kami\'{n}ska, Physica  {\bf A 365}, 360-382, (2006)
\bibitem{broglie} L. de Broglie, {\it La Thermodynamique de la particule isol\'{e}e}, Gauthier-Villars, Paris, 1964
\bibitem{gar3} P. Garbaczewski, Phys. Rev.  {\bf E 59}, (1999), 1498--1511
\end{thebibliography}
\end{document}